\documentclass[aps,prd,showpacs,bibnotes,amssymb,amsfonts,footinbib,showkeys,amsmath]{revtex4}
\usepackage{axodraw}
\usepackage{graphicx}
\usepackage{dcolumn}
\usepackage{bm}
\begin{document}
\title{\Large\bf Supergravity from Bosonic String}
\medskip
\preprint{}
\author{\bf B. B. Deo}
\affiliation{Department of Physics, Utkal University, Bhubaneswar, India}
\begin{abstract}
Deser-Zumino~\cite{Deser76} N=1 supergravity action for the graviton and gravitino pair in four 
dimensions is deduced from the Nambu-Goto action~\cite{Nambu70,Goto71} of the 26 dimensional
 bosonic string theory using Principle of Equivalence.
\end{abstract}
\pacs{04.65.+e, 12.60.J}
\keywords{supergravity, supersymmetry, superstring}
\date{\today}
\maketitle

Not long ago, Casher,Englert, Nicolai and Taormina ~\cite{Casher85,Englert01} showed that consistent 
superstrings can be the solutions of D=26 bosonic string theory and the later appears as the fundamental
string theory. This has also been demonstrated by me and in collaboration with L. Maharana in recent
references~\cite{Deo103} and ~\cite{Deo04,Deo02,Deo203}. Details of constructon of consistent 
superstring are given there.
None of the existing four dimensional string theories contain  construction of gravitino and graviton ground
states from string theory excitation quanta leading to supergravity. As far as I know, this is the first
attempt in this direction.
I repeat some of them so that the preprints can be skipped without missing the deductions of the main theme
which is of immediate information to string theorists.
The physical open bosonic string theory with the coordinates $X^{\mu}$ in the world sheet $(\sigma,\tau)$
is given by the action

\begin{equation}
S=-\frac{1}{2\pi}~\int d^2\sigma~\partial^{\alpha}X^{\mu}(\sigma,\tau)~
\partial_{\alpha}X_{\mu}(\sigma,\tau),~~~~~~~~~~~~~~~~~\mu=0,1,2..........,25.
\label{eq1}
\end{equation}
 It is well known that we can introduce 44 Majorana fermions $\phi_{\lambda}$
as proven basically by Mandelstam~\cite{Mandelstam75} for (1+1) quantum field theory, which are scalars 
in the world sheet and four bosonic coordinates and the action
\begin{equation}
 S=-\frac{1}{2\pi}~\int~d^2 \sigma \left [ \partial^{\alpha}X^{\mu}(\sigma,\tau)~
\partial_{\alpha}X_{\mu}(\sigma,\tau) - i \sum_{j=1}^{44}~\bar{\phi}^j \rho^{\alpha}
\partial _{\alpha}\phi_j\right ], ~~~~\mu=0,1,2,3.\label{eq2}
\end{equation}
where we have taken 
\begin{eqnarray}
\partial_{\alpha}=( \partial_{\sigma},\partial_{\tau}),~~~~~~~
\rho^0 =
\left (
\begin{array}{cc}
0 & -i\\
i & 0\\
\end{array}
\right ),~~~~~~~~~~
\rho^1=
\left (
\begin{array}{cc}
0 & i\\
i& 0\\
\end{array}
\right )~~~~ \text{and}~~~~
\bar{\phi}=\phi^{\dag}\rho^0\label{eq3}
\end{eqnarray}
In the Chapter-6.1.2 in \cite{Green87}, it has also been proved and stated that two Majorana fermions
in 1+1 dimensions are equivalent to one real boson in the infinite volume limit. It is also true
in anomaly free string theory on a finite interval or a circle even one like in Veneziano model.
Observing that there are also Majorana Lorentz fermions $\psi^{\mu}$ in the bosonic representation 
0f SO(3,1), we can further rewrite equation (\ref{eq2}) as~\cite{Deo103}
\begin{equation}
S=-\frac{1}{2\pi}~\int~d^2 \sigma \left [ \partial^{\alpha}X^{\mu}(\sigma,\tau)~
\partial_{\alpha}X_{\mu}(\sigma,\tau) - i \sum_{j=1}^{11}~\bar{\psi}^{\mu,j} \rho^{\alpha}
\partial _{\alpha}\psi_{\mu,j}\right ]\label{eq4}
\end{equation}
presuming this to be anomaly free. This has been proved in reference~\cite{Deo103}.
The upper index refer to row and the lower index to a column. Searching for a $\Psi^{\mu}$, to copartner to
$X^{\mu}$, one has to further divide the elevens into two groups. There is a group of six, j=1,2,.,6, 
where the negative and positive frequency parts are
$\psi^{\mu,j}= \psi^{(+)\mu,j}+ \psi^{(-)\mu,j}$ and another group of five, k=7,8,..11 with
$\phi^{\mu,k}=\phi^{+\mu,k}- \phi^{-\mu,k}$. The negative sign is absorbed in creation operators as allowed
by the freedom of phase of Majorana fermions. I introduce matrices $(e^j, e^k)$ which are rows of ten 
zeroes and only one 1 in the $j^{th}$ or $k^{th}$ place. $e^je_j$=6 and $e^ke_k$=5. With these the 
world sheet action,
\begin{equation}
S=-\frac{1}{2\pi}~\int~d^2 \sigma \left [ \partial^{\alpha}X^{\mu}(\sigma,\tau)~
\partial_{\alpha}X_{\mu}(\sigma,\tau) - i\bar{\psi}^{\mu,j} \rho^{\alpha}
\partial _{\alpha}\psi_{\mu,j} + i\bar{\phi}^{\mu,k} \rho^{\alpha}
\partial _{\alpha}\phi_{\mu,k}\right ]\label{eq5}
\end{equation}
is invariant under supersymmetric transformations,
\begin{eqnarray}
\delta X^{\mu} &=& \bar{\epsilon} \left (e^j\psi^{\mu}_j - e^k\phi_k^{\mu}\right ),\label{eq7}\\
\delta\psi^{\mu,j}&=& -i\epsilon~ e^j\rho^{\alpha}\partial_{\alpha} X^{\mu},\label{eq8}\\
\text{and} ~~~~~~~~\delta\phi^{\mu,k}&=& -i\epsilon~ e^k\rho^{\alpha}\partial_{\alpha} X^{\mu}.\label{eq9}
\end{eqnarray}
$\epsilon$ is a constant anticommuting spinor.

In action (\ref{eq5}), there are 4 bosons but eleven fermions which are Lorentz vectors of SO(3,1) which
is possible only in 1+1 dimensions~\cite{Green87}. 
This will show up in two supersymmetric successive transformations. In our case, in particular, 
the transformation leads to spatial translation with the
coefficient $a^{\alpha}= 2i\bar{\epsilon}^1\rho^{\alpha}\epsilon_2$ but constrained by 
\begin{equation}
\psi^{\mu}_j= e_j\Psi^{\mu}, ~~~~~~~~~\phi^{\mu}_k= e_k\Psi^{\mu}\label{eq13}
\end{equation}
which defines the nature of $e_j$ and $e_k$ and also
provides the much needed linear combination for the fermion SUSY partner of $X^{\mu}$
\begin{equation}
\Psi^{\mu}=e^j\psi^{\mu}_j - e^k\phi^{\mu}_k .\label{eq14}
\end{equation}
There is no need for additional auxilliary field.
The action (\ref{eq5}) reduces to
\begin{equation}
S=-\frac{1}{2\pi}~\int~d^2 \sigma \left [ \partial^{\alpha}X^{\mu}(\sigma,\tau)~
\partial_{\alpha}X_{\mu}(\sigma,\tau) - i\bar{\Psi}^{\mu} \rho^{\alpha}
\partial _{\alpha}\Psi_{\mu} \right ]\label{eq17}
\end{equation}
Using equations (\ref{eq7}) to (\ref{eq9}) and equation (\ref{eq17}) itself, it is easy to verify 
\begin{eqnarray}
\delta X^{\mu}&=& \bar{\epsilon}\Psi^{\mu},~~~~~~~\delta \Psi^{\mu} = -ie \rho^{\alpha}\partial_{\alpha}X^{\mu}
\label{eq15}\\
\left [\delta_1, \delta_2\right ]X^{\mu}&=& a^{\alpha}\partial_{\alpha}X^{\mu},~~~~~~~~~~
\left [\delta_1, \delta_2\right ]\Psi^{\mu,j} = a^{\alpha}\partial_{\alpha}\Psi^{\mu,j}.\label{eq16}
\end{eqnarray}
The actions given in equations(\ref{eq5}) and (\ref{eq16}) are equivalent. $\Psi^{\mu}$, 
while in j$^{th}$ or k$^{th}$site emits and absorbs quanta by quantising $\psi^{\mu, j}$ or $\phi^{\mu, k}$
according to the equation (\ref{eq5}) which should be unitary, free of ghosts and nomalies. It should be
free of tachyons and modular invariant. These aspects have been studied in detail in references~\cite{Deo203}
and ~\cite{Deo04}, where, also currents and energy momentum tensors have been given in light cone basis
as deduced from local 2d-global symmetric extension of the above action. We shall not need these details here.

The action in (\ref{eq17}) yields the only nonvanishing equal time commutator and anticommutators~\cite{Green87}
as
\begin{eqnarray}
\left [ \partial_{\pm}X^{\mu}(\sigma, \tau), \partial_{\pm}X^{\nu}(\sigma', \tau)\right ]&=& \pm\frac{\pi}{2}
\eta^{\mu\nu}\delta'(\sigma-\sigma')\label{eq18}\\
\{ \psi_A^{\mu}(\sigma,\tau), \psi_B^{\nu}(\sigma',\tau)\}&=&\pi \eta^{\mu\nu}\delta(\sigma-\sigma')\delta_{AB}
\label{eq19}\\
\{ \phi_A^{\mu}(\sigma,\tau), \phi_B^{\nu}(\sigma',\tau)\}&=&-\pi \eta^{\mu\nu}\delta(\sigma-\sigma')\delta_{AB}
\label{eq20}\\
\{ \Psi^{\mu}_A(\sigma,\tau), \Psi^{\nu}_B(\sigma',\tau)\}&=&\pi \eta^{\mu\nu}\delta(\sigma-\sigma')\delta_{AB}
\label{eq21}
\end{eqnarray}
On a light cone basis, the Noether supercurrents and energy momentum tensors are
\begin{eqnarray}
J_+&=&\partial_+X_{\mu}\Psi_+^{\mu}\label{eq23}\\
J_-&=&\partial_-X_{\mu}\Psi_-^{\mu}\label{eq24}\\
T_{++}&=&\partial_+X^{\mu}\partial_+X_{\mu} +\frac{i}{2}\bar{\psi}_+^{\mu,j}
\partial_+\psi_{+\mu,j} -\frac{i}{2}\bar{\phi}_+^{\mu,k}
\partial_+\phi_{+\mu,k} \label{eq25}\\
T_{--}&=&\partial_-X^{\mu}\partial_-X_{\mu} +\frac{i}{2}\bar{\psi}_-^{\mu,j}
\partial_-\psi_{-\mu,j} -\frac{i}{2}\bar{\phi}_-^{\mu,k}
\partial_-\phi_{-\mu,k}\label{eq26}
\end{eqnarray}
It follows from the variation of the local 2-d supersymmetric action that the above equations (\ref{eq23}) to
(\ref{eq26}) vanish. This enables one to construct physical states without ghosts in the Fock space.
One quantises the action (\ref{eq5}) as ~\cite{Deo103}
\begin{eqnarray}
X^{\mu}(\sigma,\tau)&=& x^{\mu} +p^{\mu}\tau + i\sum_{n\neq0}\frac{1}{n} \alpha_n^{\mu}~e^{-in\tau}~Cos(n\sigma)
\label{eq51}\\
\text{or}~~~~~~~~~\partial_{\pm} X^{\mu} &=& \frac{1}{2}\sum_{-\infty}^{+\infty}\alpha^{\mu}_n~
e^{-in(\tau\pm \sigma)}\label{eq52}\\
\text{and}~~~~~~~~\left [ ~\alpha^{\mu}_m, ~\alpha_n^{\nu}~\right ]&=& m~\delta_{m+n}~\eta^{\mu\nu}\label{eq53}
\end{eqnarray}
With Neveu-Schwartz(NS) and Ramond (R)~\cite{Neveu71,Ramond71} boundary conditions
\begin{eqnarray}
\Psi^{\mu}_{\pm}&=&\frac{1}{\sqrt{2}}\sum_{r\in Z+\frac{1}{2}} B^{\mu}_r ~e^{-ir(\tau\pm\sigma)},~~
\text{with}~~~ b_r^{\mu,j}=e^jB_r^{\mu}~~\text{and}~~b^{'\mu,k}=e^kB_r^{\mu} ~~~~~~~~\text{NS}
\label{eq171}\\
&=&\frac{1}{\sqrt{2}}\sum_m D_m^{\mu}~e^{-im(\tau\pm\sigma)},~~\text{with}~~ d_m^{\mu,j}=e^jB^{\mu}_m,~~\text{and}
~~d_m^{'\mu,k}=e^kB_m^{\mu} ~~~~~~~\text{R}\label{eq172}
\end{eqnarray}
and the anticommutation relations are
\begin{equation}
\{ B^{\mu}_r,~B_s^{\nu}\} =\eta^{\mu\nu}\delta_{r+s},~~~~~~~\{ D^{\mu}_m,~D_n^{\nu}\} =
\eta^{\mu\nu}\delta_{m+n}.\label{eq173}
\end{equation}
To construct admissible states, it is essential to give the Super Virasoro generators and their algebra
in notations of reference ~\cite{Neveu71,Ramond71}
\begin{eqnarray}
L_m&=&\frac{1}{\pi}\int_{-\pi}^{\pi}d\sigma e^{im\sigma}T_{++}\label{eq62}\\
 &= &\frac{1}{2}\sum^{\infty}_{-\infty}:\alpha_{-n}\cdot\alpha_{m+n}: +\frac{1}{2}
\sum_{r\in z+\frac{1}{2}}(r+\frac{1}{2}m): (b_{-r} \cdot b_{m+r} - b_{-r}' \cdot b_{m+r}'):~~~\text{NS}
\label{eq63}\\
&=&\frac{1}{2}\sum^{\infty}_{-\infty}:\alpha_{-n}\cdot\alpha_{m+n}: +\frac{1}{2}
\sum^{\infty}_{n=-\infty}(n+\frac{1}{2}m): (d_{-n} \cdot d_{m+n} - d_{-n}'
\cdot d_{m+n}'),: ~~~\text{R}\label{eq64}\\
G_r &=&\frac{\sqrt{2}}{\pi}\int_{-\pi}^{\pi}d\sigma e^{ir\sigma}J_{+}=
\sum_{n=-\infty}^{\infty}\alpha_{-n}\cdot \left( e^jb_{n+r,j}- e^kb'_{n+r,k}\right ), 
~~~~~~\text{NS}\label{eq65}\\
F_m &=&\sum_{-\infty}^{\infty} \alpha_{-n}\cdot \left( e^jb_{n+r,j}- e^kb'_{n+r,k}\right ) .
~~~~\text{R}\label{eq66}
\end{eqnarray}
The Super Virasoro algebra is
\begin{eqnarray}
\left [L_m , L_n\right ] & = &(m-n)L_{m+n} +\frac{C}{12}(m^3-m)\delta_{m,-n},\label{eq68}\\
\left [L_m , G_r\right ] & = &(\frac{1}{2}m-r)G_{m+r}, ~~~~~~~~~~~~~~~~~~~~~~~~~~~~~~\text{NS}\label{eq69}\\
\{G_r , G_s\} & =& 2L_{s+r} +\frac{C}{3}(r^2-\frac{1}{4})\delta_{r,-s},\label{eq70}\\
\left [L_m , F_n\right ] & = & (\frac{1}{2}m-n)F_{m+n},~~~~~~~~~~~~~~~~~~~~~~~~~\hfill \text{R}\label{eq71}\\
\{F_m, F_n\} & = & 2L_{m+n} +\frac{C}{3}(m^2-1)\delta_{m,-n},\;\;\;\; m\neq 0.\label{eq72}
\end{eqnarray}
The central charge C of the action (\ref{eq5}) is 26.

In this superstring theory, there are four bosons (C=4), twenty two transverse fermions (C=11)
($\psi^{1,2}_j, \phi^{1,2}_k$) belonging to two transverse superfermionic modes $\Psi^{1,2}$ (equation 9),
the usual conformal ghosts \cite{Green87,Kaku99} (C=-26) and lastly the superconformal ghosts ($\beta,\gamma$)
 contributing a central charge 11 in a supplemented Hilbert space. The last action is identical to the
action of the 22 longitudinal ghost modes ($\psi^{0,3}_j, \phi^{0,k}_k$) or two superfermionic 
modes $\Psi^{0,3}$. This identity has been proved in the reference [9]. Thus the central charge is zero
and there is Lorentz invariance and the superstring theory is anomaly free.

The physical states are defined as
\begin{eqnarray}
\text{NS:} ~~~~~~~(L_o-1)|\phi>&=&0, ~~~L_m|\phi>=0, ~~~G_r|\phi>= 0.~~~~~ \text{r,m}~ > ~~0.\label{eq81}\\
\text{R:} ~~~~~~~(L_o-1)|\psi>_{\alpha}&=&(F^2_o-1)|\psi>_{\alpha}=0, ~~~L_m|\psi>_{\alpha}=0, 
~~~F_r|\psi>=0.~~~~~ \text{m}~ > ~~0.\label{eq81a}
\end{eqnarray}
$\alpha$ is the spinor component index. Admissible Fock space states are
\begin{eqnarray}
\text{NS}~~~ \text{Eigenstates} &:&~~~\prod_{n,\mu}\prod_{m,\nu}\left \{\alpha_{-n}^{\mu}\right \}
\left \{ B^{\nu}_{-m}
\right \}|0>,\label{eq174}\\
\text{R}~~~~ \text{Eigenstates} &:&~~~\prod_{n,\mu}\prod_{m,\nu}\left \{\alpha_{-n}^{\mu}\right \}
\left \{ D^{\nu}_{-m}\right \}|0>u.\label{eq175}
\end{eqnarray}
G.S.O. projection operators $(1+(-1)^F)$ ~\cite{Green87} is implicit for the NS eigenstates.
Supersymmetry forbids the existence of tachyonic vacuum. Here, there are two vacua which are 
tachyonic in the scalar NS sector and the other in the fermionic Ramond sector. Their self energies cancel
 as in the supersymmetry, not showing up in overall Fock space.

In the Ramond or Fermionic sector, I aim at the $\frac{3}{2}$-spin gravitino. In four flat dimensions, the 
Rarita-Schwinger equation is 
\begin{equation}
\epsilon^{\mu\nu\lambda\sigma}\gamma_5\gamma_{\lambda}\partial_{\nu}\psi_{\sigma}=  
\left ( g^{\mu\nu}\gamma^{\sigma} - g^{\mu\sigma}\gamma^{\nu}+g^{\nu\sigma}\gamma^{\mu}-
\gamma^{\mu}\gamma^{\nu}\gamma^{\sigma}\right )\partial_{\nu}\psi_{\sigma}=0\label{eq188}
\end{equation}
As shown by GSO in reference ~\cite{Gliozzi77}, this is equivalent to
\begin{equation}
p^{\mu}\psi_{\mu}=\gamma^{\mu}\psi_{\mu}=\gamma\cdotp\psi_{\mu}=0\label{eq189}
\end{equation}
in off-shell momentum space. Condition (\ref{eq189}) ensures that there is no admixture of spin $\frac{1}{2}$
fermions.

Let the Fock ground state above the spurious tachyonic state of the  R-sector be
\begin{equation}
|\phi_o>=\alpha_{-1}^{\mu}|0,p> u_{\mu} \text{or}   D_{-1}^{\mu}|0,p> u_{1\mu};
\end{equation}
u$_{\mu}$, $\text{u}_{1\mu}$ are spinors. This is possible because there is Ramond tachyon along with 
Neveu-Schwartz.
In either case $L_0|\phi_0>=0$, so that this state is massless.
Constraints $L_1|\phi_0>$=0 and $F_1|\phi_o>$=0 imply that 
\begin{equation}
p^{\mu}~u_{\mu}=0,~~~~~~~~\gamma^{\mu}~u_{\mu}=0
\end{equation}
It is known that $D_o^{\mu}\sim \gamma^{\mu}$, the Dirac gamma matrices and $\alpha_0^{\mu}=p^{\mu}$. So
$p^{\mu}u_{1\mu}$=0  and $\gamma^{\mu}u_{1\mu}$=0 as well. There is the freedom to choose a four component
spinor $\psi^{\mu}$ such that $u_{1\mu}=\gamma^5 u_{\mu}$ and
\begin{equation}
\psi_{\mu}=(1+\gamma_5)u_{\mu}
\end{equation}
By virtue of the above equations
\begin{equation}
\gamma^{\mu}\psi_{\mu}=p^{\mu}\psi_{\mu}=0
\end{equation}
Further, $F_0$ is essentially $\gamma\cdot p$. The wave equation is 
\begin{equation}
F_0\psi_{\mu}\sim \gamma\cdot p\psi_{\mu}=0
\end{equation}
So $\psi_{\mu}$ is the Rarita-Schwinger spin $\frac{3}{2}$ vectorial spinor without spin $\frac{1}{2}$
component. The general relativity equation can be found by invoking the principle of equivalence. The 
every curved space time point on the manifold, a tangent space is constructed. The vector indices are a,b,.. 
The vierbein matrices like $e^a_{\mu}$ takes the x space objects to the tangent space and vice versa
\begin{equation}
e^a_{\mu}e^b_{\nu}=g^{\mu\nu},~~~~~~~~~~~~e^a_{\mu}e^{b\mu}=\delta^{ab}
\end{equation}
For completely
specifying the spin connections $\omega_{\mu}^{ab}$, the covariant derivative $D_{\mu}$ is such that

\begin{equation}
D_{\mu}e^a_{\lambda}=0
\end{equation}
With defination of $\gamma$ matrices
\begin{equation}
\gamma^ae_a^{\mu}=\gamma^{\mu},
\end{equation}
the covariant derivative of a spinor $\psi$ is
\begin{equation}
D_{\mu}\psi =(\partial_{\mu} +\omega_{\mu}^{ab}\sigma_{ab})\psi
\end{equation}
$\sigma^{ab}$ is the antisymmetric product of two gamma matrices. Basically, the pricniple of equivalence 
states that in the curved space time, the Rarita-Schwinger equation is
\begin{equation}
\gamma_{\nu}\gamma^5 D_{\sigma}\psi_{\rho} \epsilon^{\mu\nu\sigma\rho}=0
\end{equation}

With e=$\sqrt{g}$, the invariant action in the curved space time is
\begin{equation}
S_{gravitino}=-\frac{i}{2}\int ~d^4\text{x}~\text{e}~\bar{\psi}_{\mu}\gamma_{\nu}\gamma^5D_{\sigma}\psi_{\rho}
~\epsilon^{\mu\nu\sigma\rho}\label{eq189x}
\end{equation}

For the graviton, the problem is more involved but interesting. Let $C^{ij}$ be the coefficients 
of the probabilities of emission and absorption of the superpartners $\Psi^{\mu}$ in the $i^{th}$ and
$j^{th}$ sites at the same instance. Then consider the string state(dropping the creation operator suffix `-1/2')
above the spurious NS tachyonic state,

\begin{equation}
h^{\dag\mu\nu}(p)=\sum_{i,j}C^{ij}\left ( b_i^{\mu} b_j^{\nu}+ b_i^{\nu} b_j^{\mu}-2\eta^{\mu\nu}
 b_i^{\lambda} b_j^{\lambda}\right )|0,p>.\label{eq189a}
\end{equation}
The presence of NS tachyon help build the massless graviton state. This is symmetric and traceless. 
Further, $L_0$ will be taken as the free Hamiltonian $H_o$ in the interaction representation. 
Operating on this state, one gets $L_0 h^{\dag\mu\nu}(p)=0$. So this is massless. Also 
$p_{\mu}h^{\dag\mu\nu}(p)=p_{\nu}h^{\dag\mu\nu}(p)=0 $ if $C^{ij}=C^{ji}$. If $C^{ij}$ is symmetric,
a little algebra shows that
\begin{equation}
G_{\frac{1}{2}}h^{\dag}_{\mu\nu}(p)=\left[G_{\frac{1}{2}}, ~h^{\dag}_{\mu\nu}(p) \right ]=0.\label{eq189b}
\end{equation}
Thus $h^{\dag}_{\mu\nu}(p)$ satisfy all physical state conditions due to Virasoro algebraic relations.
This should be graviton. The commutator is
\begin{equation}
\left [h_{\mu\nu}(p), ~h_{\mu\nu}(q)\right ]=2\pi|C|^2\delta^4(p-q).
\end{equation}
To switch over quantum field theory, we define the gravitational field by space time Fourier transform of
$h_{\mu\nu}(p)$
\begin{equation}
h_{\mu\nu}(x)=\frac{1}{(2\pi)^3}\int \frac{d^3p}{\sqrt{2p_0}}
\left [h_{\mu\nu}(p)~e^{ipx}+h^{\dag}_{\mu\nu}(p)~e^{-ipx}\right ],
\end{equation}
with the commutator
\begin{equation}
\left [h_{\mu\nu}(x), ~h_{\lambda\sigma}(y)\right ]=\frac{1}{(2\pi)^3}\int \frac{d^3p}{2p_0}
\left [ e^{ip.(x-y)} - e^{-ip.(x-y)}\right ]f^{\mu\nu,\lambda\sigma},
\end{equation}
where
\begin{equation}
f^{\mu\nu,\lambda\sigma}=g^{\mu\lambda}g^{\nu\sigma}+g^{\nu\lambda}g^{\mu\sigma}-g^{\mu\nu}g^{\sigma\lambda}.
\end{equation}
The Feynman propagator is
\begin{equation}
\triangle^{\mu\nu,\lambda\sigma}(x-y)=<0|T~h_{\mu\nu}(x)h_{\lambda\sigma}(y)|0>=\frac{1}{(2\pi)^4}
\int d^4p~ \triangle^{\mu\nu,\lambda\sigma}(p)~e^{ip.(x-y)},
\end{equation}
where
\begin{equation}
\triangle^{\mu\nu,\lambda\sigma}(p)=\frac{1}{2}~f^{\mu\nu,\lambda\sigma}\frac{1}{p^2-i\epsilon}.
\end{equation}
This is the propagator of the graviton in the interaction representation. But we do not yet have the
general relativity. I take the clue from the case of gravitino and follows it up. In four dimensions,
the fourier transform is
\begin{equation}
h_{\mu\nu}(x)=\int \frac{d^4p}{(2\pi)^4} ~h_{\mu\nu}(p)~e^{ipx}
\end{equation}
and the hermitian
\begin{equation}
h^{\dag}_{\mu\nu}(x)=\int \frac{d^4p}{(2\pi)^4} ~h^{\dag}_{\mu\nu}(p)~e^{-ipx}
\end{equation}

They are related by
\begin{equation}
h^{\dag\mu}_{\lambda}h^{\lambda}_{\nu}=\delta_{\mu\nu}\label{eq189c}
\end{equation}
by choosing the normalisation
\begin{equation}
<0,p|0,q>=(2\pi)^4\delta^4(p-q)
\end{equation}
and absorbing $\delta^{(4)}(0)$ in $|C|^2$. In tangent space,
\begin{equation}
h_{\mu\nu}(x)=e^a_{\mu}(x)e^b_{\nu}(x)T_{ab}
\end{equation}
with $h_{\mu\mu}(x)=T_{ab}$=0. After some calculation, one gets

\begin{equation}
h_{\mu\nu}(x)=e^a_{\mu}e^b_{\nu}(x)T_{ab},~~~\text{with}~~~ h_{\mu\mu}(x)=T_{bb}=0.\label{eq190}
\end{equation}

After some calculation, one gets by equation(\ref{eq190})
\begin{equation}
\left [ D_{\mu}, D_{\lambda} \right ] h^{\lambda}_{\nu} =e_{a}^{\lambda}~e_{\nu b}~R^{ac}_{\mu\lambda}T^{cb}=0
\label{eq190c}
\end{equation}
where the Riemannian Curvature tensor is
\begin{equation}
R^{ac}_{\mu\lambda}=\partial_{\mu}\omega_{\lambda}^{ac}+\partial_{\mu}\omega_{\lambda}^{dc}-(\mu \leftrightarrow
\lambda)
\end{equation}
Inverting equation(\ref{eq190c})
\begin{equation}
\left [ D_{\mu}, D_{\lambda} \right ] h^{\lambda}_{\nu} =R_{\mu\lambda}h^{\lambda}_{\nu}=0
\label{eq190d}
\end{equation}
as the parallel transport relation.
Using the equation (\ref{eq189c}), one gets
\begin{equation}
R_{\mu\nu}=h^{\dag\lambda}_{\nu} R_{\mu\sigma}h^{\sigma}_{\lambda}=
h^{\dag\lambda}_{\nu}  \left [ D_{\mu}, D_{\sigma} \right ]  h^{\sigma}_{\lambda}\label{eq195}
\end{equation}
At this point, it is appropriate to quote the words of Misner, Thorne and Wheeler~\cite{Misner70}
'The Laws of Physics written in abstract geometric form, differ in no way whatsoever between 
curved space time and flat space time; this is guaranteed by and in fact is a mere rewording of the equivalence
principle. In equation (\ref{eq195}), the right hand side $\left [ D_{\mu}, D_{\sigma} \right ]=\left [
\partial_{\mu}, \partial_{\nu}\right ]=0$ in flat space time. So anywhere in the universe with curved
or flat space time
\begin{equation}
R_{\mu\nu}=0
\end{equation}
This is the Einstein equation of General Relativity deduced from string theory.
The graviton action is, with $R=g^{\mu\nu}R_{\mu\nu}$, 
\begin{equation}
S_{gravitation}= -\frac{1}{2\kappa}\int\text{d}^4\text{x}~e~R\label{eq191}
\end{equation}
$\kappa$ is the gravitational coupling constant, whose square is proportional to the Newton's constant.
The N=1 supergravity action in four dimensions is the sum of equations (\ref{eq189}) and (\ref{eq191}),
\begin{equation}
S=-\frac{1}{2}\int\text{d}^4\text{x}~e~\left ( \frac{1}{\kappa^2} R +\bar{\psi}_{\mu}\gamma_{\nu}
\gamma^5D_{\sigma}\psi_{\rho}~\epsilon^{\mu\nu\sigma\rho}\right )\label{eq192}
\end{equation}
The local supergravity transformations that leave it invariant are
\begin{equation}
\delta\psi_{\mu}= \frac{1}{\kappa} D_{\mu}~\epsilon(x)~~~~~~~ \text{and} ~~~~~~\delta~e_{\mu}^m=
-\frac{i}{2}\kappa ~\bar{\epsilon}(x)\gamma^m\psi_{\mu}
\end{equation}

I have succeded in deriving the 4d, N=1 supergravity theory as it develops from a renormalisable open
26d bosonic string. This has unlimited useful consequences in physics.

I thank Prof L. Maharana but for whose discussions and suggestions, this article could not have been
completed.

\end{document}